# The criterion for maximally entangled four-qubit state

Xin-wei Zha[*], Hai-yang Song, Feng Feng

School of Science, Xi'an University of Posts and Telecommunications, Xi'an, 710121, P R China

**Abstract**   Paolo Facchi et al.[Phys. Rev. A. 77, 060304( R) (2009)] presented a maximally multipartite entangled state(MMES). Here we give the criterion for four-qubit state.and some new characterizations of the maximally entangled four-qubit state are given. .

Quantum entanglement is a valuable resource for the implementation of quantum computation and quantum communication protocols, like quantum teleportation [1], remote state preparation [2-4] and so on. Due to its great relevance, from both the theoretical and practical points of view,it is imperative to explore and characterize all aspects of the quantum entanglement of multipartite quantum systems. A considerable amount of research has been devoted to the study of multi-qubit entanglement measures defined as the sum of bipartite entanglement measures over all the possible bi-partitions of the full system[5-19].Recently, Paolo Facchi et al [5] defined a MMES (maximally multipartite entangled state) as a minimizer of what they called the potential of multipartite entanglement. namely

$$\pi_{ME} = \binom{n}{n_A}^{-1} \sum_{|A|=n_A} \pi_A, \qquad (1)$$

where $n_A = [n/2]$, and

$$\pi_A = Tr_A \rho_A^2, \quad \rho_A = Tr_{\bar{A}} |\psi\rangle\langle\psi|. \qquad (2)$$

Here $Tr_X$ denotes the partial trace over subsystem X. For n=4 qubits, they numerically obtained $\min \pi_{ME} \simeq 0.333$. This is consistent with results obtained by other authors[6,7]. The quantity $\pi_{ME}$ measures the average bipartite entanglement over all possible balanced bipartition. For four qubits, we have

$$\pi_{ME} = \frac{1}{6}(\pi_{12} + \pi_{13} + \pi_{14} + \pi_{23} + \pi_{24} + \pi_{34}), \qquad (3)$$

---

where $\pi_{12} = Tr_{12}\rho_{12}^2$, $\rho_{12} = Tr_{34}|\varphi\rangle_{1234\ 1234}\langle\varphi|$ and so on.

Without any loss of generality, we first consider a general four-qubit state

$$\begin{aligned}|\varphi\rangle_{1234} = (&a_0|0000\rangle + a_1|0001\rangle + a_2|0010\rangle + a_3|0011\rangle \\ &+ a_4|0100\rangle + a_5|0101\rangle + a_6|0110\rangle + a_7|0111\rangle \\ &+ a_8|1000\rangle + a_9|1001\rangle + a_{10}|1010\rangle + a_{11}|1011\rangle \\ &+ a_{12}|1100\rangle + a_{13}|1101\rangle + a_{14}|1110\rangle + a_{15}|1111\rangle)_{1234}\end{aligned} \quad (4)$$

and it is assumed that the wave function satisfies the normalization condition $\sum_{i=0}^{15}|a_i|^2 = 1$.

From Eqs.(2) and (4), we can obtain

$$\begin{aligned}\pi_{12} &= Tr_{12}\rho_{12}^2 \\ &= \left(|a_0|^2 + |a_1|^2 + |a_2|^2 + |a_3|^2\right)^2 + \left(|a_4|^2 + |a_5|^2 + |a_6|^2 + |a_7|^2\right)^2 \\ &\quad + \left(|a_8|^2 + |a_9|^2 + |a_{10}|^2 + |a_{11}|^2\right)^2 + \left(|a_{12}|^2 + |a_{13}|^2 + |a_{14}|^2 + |a_{15}|^2\right)^2 \\ &\quad + 2|a_0 a_4^* + a_1 a_5^* + a_2 a_6^* + a_3 a_7^*|^2 + 2|a_0 a_8^* + a_1 a_9^* + a_2 a_{10}^* + a_3 a_{11}^*|^2 \\ &\quad + 2|a_0 a_{12}^* + a_1 a_{13}^* + a_2 a_{14}^* + a_3 a_{15}^*|^2 + 2|a_4 a_8^* + a_5 a_9^* + a_6 a_{10}^* + a_7 a_{11}^*|^2 \\ &\quad + 2|a_4 a_{12}^* + a_5 a_{13}^* + a_6 a_{14}^* + a_7 a_{15}^*|^2 + 2|a_8 a_{12}^* + a_9 a_{13}^* + a_{10} a_{14}^* + a_{11} a_{15}^*|^2\end{aligned} \quad (5)$$

Similarly, we can obtain $\pi_{13}, \pi_{14}\cdots$, and $\pi_{34}$ (see Appendix A). Then we have

$$\pi_{ME} = \frac{1}{6}(\pi_{12} + \pi_{13} + \pi_{14} + \pi_{23} + \pi_{24} + \pi_{34}) = \frac{1}{3}(1+K), \quad (6)$$

and $K = K_1 + K_2 \geq 0$ (see Appendix B).

Hence, for maximally entangled four qubit states, it must be $K_1 + K_2 = 0$.

To show the applicability of this criterion, we shall examine some known examples. Let us first consider the following state

$$|\varphi\rangle_{1234} = (a_3|0011\rangle + a_5|0101\rangle + a_6|0110\rangle + a_9|1001\rangle + a_{10}|1010\rangle + a_{12}|1100\rangle)_{1234} \quad (7)$$

According to Appendix B, if $a_3 = a_5 = a_6 = a_9 = a_{10} = a_{12} = \frac{1}{\sqrt{6}}$, we have $K = 1$;

if $a_3 = a_{12} = \frac{1}{\sqrt{6}}, a_5 = a_{10} = \frac{\omega}{\sqrt{6}}$, and $a_6 = a_9 = \frac{\omega^2}{\sqrt{6}}$, with $\omega = \exp(2i\pi/3)$, then $K = 0$.

This state was discovered by Higuchi and Sudbery [6], which is given by

$$|HS\rangle_{1234} = \frac{1}{\sqrt{6}}\left[|0011\rangle + |1100\rangle + \omega(|0101\rangle + |1010\rangle) + \omega^2(|0110\rangle + |1001\rangle)\right]. \quad (8)$$

It is easy to show that $\pi_{12} = \frac{1}{3}$, $\pi_{13} = \frac{1}{3}$ and $\pi_{14} = \frac{1}{3}$; hence $\pi_{ME} = \frac{1}{3}$.

Given a state

$$\begin{aligned}|\varphi\rangle_{1234} = (&a_0|0000\rangle + a_3|0011\rangle + a_5|0101\rangle + a_6|0110\rangle \\ &+ a_9|1001\rangle + a_{10}|1010\rangle + a_{12}|1100\rangle + a_{15}|1111\rangle)_{1234}\end{aligned} \quad (9)$$

One can see that if $a_0 = a_3 = a_5 = a_6 = a_9 = a_{10} = a_{12} = a_{15} = \frac{1}{2\sqrt{2}}$, then $K = 1$;

if $a_0 = a_{15} = \frac{1}{2\sqrt{2}}$, $a_3 = a_{12} = \frac{e^{i\frac{\pi}{4}}}{2\sqrt{2}}$, $a_5 = a_{10} = \frac{e^{i\frac{\pi}{2}}}{2\sqrt{2}}$ and $a_6 = a_9 = \frac{e^{i\frac{3\pi}{4}}}{2\sqrt{2}}$, then $K = 0$;

in addition, if $a_0 = a_6 = a_9 = a_{10} = a_{12} = a_{15} = \frac{1}{2\sqrt{2}}$, $a_3 = a_5 = -\frac{1}{2\sqrt{2}}$, then $K = 0$.

This state was presented by Yeo and Chua [19], which is called the genuine four-qubit entanglement state, namely

$$|YC\rangle_{1234} = \frac{1}{2\sqrt{2}}(|0000\rangle - |0011\rangle - |0101\rangle + |0110\rangle \\ + |1001\rangle + |1010\rangle + |1100\rangle + |1111\rangle)_{1234}. \quad (10)$$

It is easy to show that $\pi_{12} = \frac{1}{4}$, $\pi_{13} = \frac{1}{4}$, $\pi_{14} = \frac{1}{2}$.

Next for the state

$$|\varphi\rangle_{1234} = (a_0|0000\rangle + a_5|0101\rangle + a_{10}|1010\rangle + a_{15}|1111\rangle)_{1234}, \quad (11)$$

if $a_0 = a_5 = a_{10} = a_{15} = \frac{1}{2}$, then $K = 1$; if $a_0 = a_{15} = \frac{1}{2}$, $a_5 = a_{10} = \frac{i}{2}$, then $K = 0$;

if, $a_0 = a_5 = a_{10} = \frac{1}{2}$, $a_{15} = -\frac{1}{2}$, $K = 0$. This state was just the cluster state given in [20], defined as

$$|\varphi\rangle_{1234} = \frac{1}{2}(|0000\rangle + |0101\rangle + |1010\rangle - |1111\rangle)_{1234}. \quad (12)$$

As the last instance, let us consider the state

$$|\varphi\rangle_{1234} = (a_0|0000\rangle + a_3|0011\rangle + a_6|0110\rangle + a_{11}|1011\rangle + a_{13}|1101\rangle + a_{14}|1110\rangle)_{1234} \quad (13)$$

Olso we can obtain, if $a_0 = a_3 = a_6 = a_{11} = a_{13} = a_{14} = \frac{1}{\sqrt{6}}$, then $K = \frac{5}{9}$;

if $a_0 = a_{13} = \frac{1}{2}$, $a_3 = a_{14} = \frac{1}{2\sqrt{2}}$ and $a_6 = a_{11} = \frac{i}{2\sqrt{2}}$, $K = 0$;

if $a_0 = a_{13} = \frac{1}{2}$, $a_3 = a_6 = a_{11} = \frac{1}{2\sqrt{2}}$ and $a_{14} = -\frac{1}{2\sqrt{2}}$, $K = 0$.

This state was discovered by Brown[8], which is given by

$$|\varphi\rangle_{1234} = \frac{1}{2}(|0000\rangle + |+011\rangle + |1101\rangle + |-110\rangle)_{1234}, \quad (14)$$

where $|+\rangle = \frac{1}{\sqrt{2}}(|0\rangle + |1\rangle)$ and $|-\rangle = \frac{1}{\sqrt{2}}(|0\rangle - |1\rangle)$ \quad (15)

In summary, we have presented a criterion for maximally entangled four-qubit states. Some new form of the maximally entangled four-qubit state are given. The method can easily extended to multi-qubit state. Brown et al.[19] conjectured that multi-qubit states of maximum entanglement always have all their single-qubit marginal density matrices completely mixed. The results obtained here is consistent with the Brown's conjecture. Borras et al.[7] conjectured that there is no pure state of seven qubits whose marginal density matrices for subsystems of 1,2,or 3 qubits are all completely mixed. We conjecture that seven and eight qubit states of maximum entanglement should have all their single-qubit and two-qubit marginal density matrices completely mixed.

Acknowledgements

This work is supported by the National Natural Science Foundation of China (Grant No. 10902083) and Shaanxi Natural Science Foundation under Contract (No. 2009JM1007).

Appendix A

Similarly to the Eq.(5), we can easily obtain

$$\pi_{13} = Tr_{13}\rho_{13}^2$$
$$= \left(|a_0|^2 + |a_1|^2 + |a_4|^2 + |a_5|^2\right)^2 + \left(|a_2|^2 + |a_3|^2 + |a_6|^2 + |a_7|^2\right)^2$$
$$+ \left(|a_8|^2 + |a_9|^2 + |a_{12}|^2 + |a_{13}|^2\right)^2 + \left(|a_{10}|^2 + |a_{11}|^2 + |a_{14}|^2 + |a_{15}|^2\right)^2$$
$$+ 2\left|a_0 a_2^* + a_1 a_3^* + a_4 a_6^* + a_5 a_7^*\right|^2 + 2\left|a_0 a_8^* + a_1 a_9^* + a_4 a_{12}^* + a_5 a_{13}^*\right|^2$$
$$+ 2\left|a_2 a_8^* + a_3 a_9^* + a_6 a_{12}^* + a_7 a_{13}^*\right|^2 + 2\left|a_2 a_{10}^* + a_3 a_{11}^* + a_6 a_{14}^* + a_7 a_{15}^*\right|^2$$
$$+ 2\left|a_0 a_{10}^* + a_1 a_{11}^* + a_4 a_{14}^* + a_5 a_{15}^*\right|^2 + 2\left|a_8 a_{10}^* + a_9 a_{11}^* + a_{12} a_{14}^* + a_{13} a_{15}^*\right|^2,$$

$$\pi_{14} = Tr_{14}\rho_{14}^2$$
$$= \left(|a_0|^2 + |a_2|^2 + |a_4|^2 + |a_6|^2\right)^2 + \left(|a_1|^2 + |a_3|^2 + |a_5|^2 + |a_7|^2\right)^2$$
$$+ \left(|a_8|^2 + |a_{10}|^2 + |a_{12}|^2 + |a_{14}|^2\right)^2 + \left(|a_9|^2 + |a_{11}|^2 + |a_{13}|^2 + |a_{15}|^2\right)^2$$
$$+ 2\left|a_0 a_1^* + a_2 a_3^* + a_4 a_5^* + a_6 a_7^*\right|^2 + 2\left|a_0 a_8^* + a_2 a_{10}^* + a_4 a_{12}^* + a_6 a_{14}^*\right|^2$$
$$+ 2\left|a_1 a_8^* + a_3 a_{10}^* + a_5 a_{12}^* + a_7 a_{14}^*\right|^2 + 2\left|a_1 a_9^* + a_3 a_{11}^* + a_5 a_{13}^* + a_7 a_{15}^*\right|^2$$
$$+ 2\left|a_0 a_9^* + a_2 a_{11}^* + a_4 a_{13}^* + a_6 a_{15}^*\right|^2 + 2\left|a_8 a_9^* + a_{10} a_{11}^* + a_{12} a_{13}^* + a_{14} a_{15}^*\right|^2,$$

$$\pi_{23} = Tr_{23}\rho_{23}^2$$
$$= \left(|a_0|^2 + |a_1|^2 + |a_8|^2 + |a_9|^2\right)^2 + \left(|a_2|^2 + |a_3|^2 + |a_{10}|^2 + |a_{11}|^2\right)^2$$
$$+ \left(|a_4|^2 + |a_5|^2 + |a_{12}|^2 + |a_{13}|^2\right)^2 + \left(|a_6|^2 + |a_7|^2 + |a_{14}|^2 + |a_{15}|^2\right)^2$$
$$+ 2\left|a_0 a_4^* + a_1 a_5^* + a_8 a_{12}^* + a_9 a_{13}^*\right|^2 + 2\left|a_0 a_2^* + a_1 a_3^* + a_8 a_{10}^* + a_9 a_{11}^*\right|^2$$
$$+ 2\left|a_2 a_4^* + a_3 a_5^* + a_{10} a_{12}^* + a_{11} a_{13}^*\right|^2 + 2\left|a_4 a_6^* + a_5 a_7^* + a_{12} a_{14}^* + a_{13} a_{15}^*\right|^2$$
$$+ 2\left|a_0 a_6^* + a_1 a_7^* + a_8 a_{14}^* + a_9 a_{15}^*\right|^2 + 2\left|a_2 a_6^* + a_3 a_7^* + a_{10} a_{14}^* + a_{11} a_{15}^*\right|^2,$$

$$\pi_{24} = Tr_{24}\rho_{24}^2$$
$$= \left(|a_0|^2 + |a_2|^2 + |a_8|^2 + |a_{10}|^2\right)^2 + \left(|a_1|^2 + |a_3|^2 + |a_9|^2 + |a_{11}|^2\right)^2$$
$$+ \left(|a_4|^2 + |a_6|^2 + |a_{12}|^2 + |a_{14}|^2\right)^2 + \left(|a_5|^2 + |a_7|^2 + |a_{13}|^2 + |a_{15}|^2\right)^2$$
$$+ 2\left|a_0 a_4^* + a_2 a_6^* + a_8 a_{12}^* + a_{10} a_{14}^*\right|^2 + 2\left|a_0 a_1^* + a_2 a_3^* + a_8 a_9^* + a_{10} a_{11}^*\right|^2$$
$$+ 2\left|a_1 a_4^* + a_3 a_6^* + a_9 a_{12}^* + a_{11} a_{14}^*\right|^2 + 2\left|a_4 a_5^* + a_6 a_7^* + a_{12} a_{13}^* + a_{14} a_{15}^*\right|^2$$
$$+ 2\left|a_0 a_5^* + a_2 a_7^* + a_8 a_{13}^* + a_{10} a_{15}^*\right|^2 + 2\left|a_1 a_5^* + a_3 a_7^* + a_9 a_{13}^* + a_{11} a_{15}^*\right|^2$$

$$\pi_{34} = Tr_{34}\rho_{34}^2$$
$$= \left(|a_0|^2 + |a_4|^2 + |a_8|^2 + |a_{12}|^2\right)^2 + \left(|a_1|^2 + |a_5|^2 + |a_9|^2 + |a_{13}|^2\right)^2$$
$$+ \left(|a_2|^2 + |a_6|^2 + |a_{10}|^2 + |a_{14}|^2\right)^2 + \left(|a_3|^2 + |a_7|^2 + |a_{11}|^2 + |a_{15}|^2\right)^2$$
$$+ 2\left|a_0 a_2^* + a_4 a_6^* + a_8 a_{10}^* + a_{12} a_{14}^*\right|^2 + 2\left|a_0 a_1^* + a_4 a_5^* + a_8 a_9^* + a_{12} a_{13}^*\right|^2$$
$$+ 2\left|a_1 a_2^* + a_5 a_6^* + a_9 a_{10}^* + a_{13} a_{14}^*\right|^2 + 2\left|a_2 a_3^* + a_6 a_7^* + a_{10} a_{11}^* + a_{14} a_{15}^*\right|^2$$
$$+ 2\left|a_0 a_3^* + a_4 a_7^* + a_8 a_{11}^* + a_{12} a_{15}^*\right|^2 + 2\left|a_1 a_3^* + a_5 a_7^* + a_9 a_{11}^* + a_{13} a_{15}^*\right|^2,$$

Appendix B

Using Appendix A and wave function normalization condition $\sum_{i=0}^{15}|a_i|^2=1$.we give

$$\pi_{ME} = \frac{1}{6}\left(\pi_{12} + \pi_{13} + \pi_{14} + \pi_{23} + \pi_{24} + \pi_{34}\right)$$

$$= \frac{1}{3}(1 + K_1 + K_2)$$

where

$$K_1 = 4\left(|a_0|^4 + |a_1|^4 + \cdots + |a_{15}|^4\right)$$
$$+ |a_0|^2 \left(2|a_1|^2 + 2|a_2|^2 - 2|a_3|^2 + 2|a_4|^2 - 2|a_5|^2 - 2|a_6|^2 - 4|a_7|^2 + 2|a_8|^2\right.$$
$$\left. - 2|a_9|^2 - 2|a_{10}|^2 - 4|a_{11}|^2 - 2|a_{12}|^2 - 4|a_{13}|^2 - 4|a_{14}|^2 - 4|a_{15}|^2\right)$$
$$+ |a_1|^2 \left(-2|a_2|^2 + 2|a_3|^2 - 2|a_4|^2 + 2|a_5|^2 - 4|a_6|^2 - 2|a_7|^2 - 2|a_8|^2\right.$$
$$\left. + 2|a_9|^2 - 4|a_{10}|^2 - 2|a_{11}|^2 - 4|a_{12}|^2 - 2|a_{13}|^2 - 4|a_{14}|^2 - 4|a_{15}|^2\right)$$
$$+ |a_2|^2 \left(2|a_3|^2 - 2|a_4|^2 - 4|a_5|^2 + 2|a_6|^2 - 2|a_7|^2 - 2|a_8|^2 - 4|a_9|^2\right.$$
$$\left. + 2|a_{10}|^2 - 2|a_{11}|^2 - 4|a_{12}|^2 - 4|a_{13}|^2 - 2|a_{14}|^2 - 4|a_{15}|^2\right)$$
$$+ |a_3|^2 \left(-4|a_4|^2 - 2|a_5|^2 - 2|a_6|^2 + 2|a_7|^2 - 4|a_8|^2 - 2|a_9|^2\right.$$
$$\left. - 2|a_{10}|^2 + 2|a_{11}|^2 - 4|a_{12}|^2 - 4|a_{13}|^2 - 4|a_{14}|^2 - 2|a_{15}|^2\right)$$
$$+ |a_4|^2 \left(2|a_5|^2 + 2|a_6|^2 - 2|a_7|^2 - 2|a_8|^2 - 4|a_9|^2 - 4|a_{10}|^2\right.$$
$$\left. - 4|a_{11}|^2 + 2|a_{12}|^2 - 2|a_{13}|^2 - 2|a_{14}|^2 - 4|a_{15}|^2\right)$$
$$+ |a_5|^2 \left(-2|a_6|^2 + 2|a_7|^2 - 4|a_8|^2 - 2|a_9|^2 - 4|a_{10}|^2\right.$$
$$\left. - 4|a_{11}|^2 - 2|a_{12}|^2 + 2|a_{13}|^2 - 4|a_{14}|^2 - 2|a_{15}|^2\right)$$
$$+ |a_6|^2 \left(2|a_7|^2 - 4|a_8|^2 - 4|a_9|^2 - 4|a_{10}|^2 - 4|a_{11}|^2\right.$$
$$\left. - 4|a_{12}|^2 - 2|a_{13}|^2 - 2|a_{14}|^2 + 2|a_{15}|^2\right)$$
$$+ |a_7|^2 \left(-4|a_8|^2 - 4|a_9|^2 - 4|a_{10}|^2 - 2|a_{11}|^2 - 4|a_{12}|^2 - 2|a_{13}|^2 - 2|a_{14}|^2 + 2|a_{15}|^2\right)$$
$$+ |a_8|^2 \left(2|a_9|^2 + 2|a_{10}|^2 - 2|a_{11}|^2 + 2|a_{12}|^2 - 2|a_{13}|^2 - 2|a_{14}|^2 - 4|a_{15}|^2\right)$$
$$+ |a_9|^2 \left(-2|a_{10}|^2 + 2|a_{11}|^2 - 2|a_{12}|^2 + 2|a_{13}|^2 - 4|a_{14}|^2 - 2|a_{15}|^2\right)$$
$$+ |a_{10}|^2 \left(2|a_{11}|^2 - 2|a_{12}|^2 - 4|a_{13}|^2 + 2|a_{14}|^2 - 2|a_{15}|^2\right)$$
$$+ |a_{11}|^2 \left(-4|a_{12}|^2 - 2|a_{13}|^2 - 2|a_{14}|^2 + 2|a_{15}|^2\right)$$
$$+ |a_{13}|^2 \left(-2|a_{14}|^2 - 2|a_{15}|^2\right) + 2|a_{14}|^2 |a_{15}|^2$$

$$\begin{aligned}
K_2 = &\ 2\left|a_0 a_4^* + a_1 a_5^* + a_2 a_6^* + a_3 a_7^*\right|^2 + 2\left|a_0 a_8^* + a_1 a_9^* + a_2 a_{10}^* + a_3 a_{11}^*\right|^2 \\
&+ 2\left|a_0 a_{12}^* + a_1 a_{13}^* + a_2 a_{14}^* + a_3 a_{15}^*\right|^2 + 2\left|a_4 a_8^* + a_5 a_9^* + a_6 a_{10}^* + a_7 a_{11}^*\right|^2 \\
&+ 2\left|a_4 a_{12}^* + a_5 a_{13}^* + a_6 a_{14}^* + a_7 a_{15}^*\right|^2 + 2\left|a_8 a_{12}^* + a_9 a_{13}^* + a_{10} a_{14}^* + a_{11} a_{15}^*\right|^2 \\
&+ 2\left|a_0 a_2^* + a_1 a_3^* + a_4 a_6^* + a_5 a_7^*\right|^2 + 2\left|a_0 a_8^* + a_1 a_9^* + a_4 a_{12}^* + a_5 a_{13}^*\right|^2 \\
&+ 2\left|a_2 a_8^* + a_3 a_9^* + a_6 a_{12}^* + a_7 a_{13}^*\right|^2 + 2\left|a_2 a_{10}^* + a_3 a_{11}^* + a_6 a_{14}^* + a_7 a_{15}^*\right|^2 \\
&+ 2\left|a_0 a_{10}^* + a_1 a_{11}^* + a_4 a_{14}^* + a_5 a_{15}^*\right|^2 + 2\left|a_8 a_{10}^* + a_9 a_{11}^* + a_{12} a_{14}^* + a_{13} a_{15}^*\right|^2 \\
&+ 2\left|a_0 a_1^* + a_2 a_3^* + a_4 a_5^* + a_6 a_7^*\right|^2 + 2\left|a_0 a_8^* + a_2 a_{10}^* + a_4 a_{12}^* + a_6 a_{14}^*\right|^2 \\
&+ 2\left|a_1 a_8^* + a_3 a_{10}^* + a_5 a_{12}^* + a_7 a_{14}^*\right|^2 + 2\left|a_1 a_9^* + a_3 a_{11}^* + a_5 a_{13}^* + a_7 a_{15}^*\right|^2 \\
&+ 2\left|a_0 a_9^* + a_2 a_{11}^* + a_4 a_{13}^* + a_6 a_{15}^*\right|^2 + 2\left|a_8 a_9^* + a_{10} a_{11}^* + a_{12} a_{13}^* + a_{14} a_{15}^*\right|^2 \\
&+ 2\left|a_0 a_4^* + a_1 a_5^* + a_8 a_{12}^* + a_9 a_{13}^*\right|^2 + 2\left|a_0 a_2^* + a_1 a_3^* + a_8 a_{10}^* + a_9 a_{11}^*\right|^2 \\
&+ 2\left|a_2 a_4^* + a_3 a_5^* + a_{10} a_{12}^* + a_{11} a_{13}^*\right|^2 + 2\left|a_4 a_6^* + a_5 a_7^* + a_{12} a_{14}^* + a_{13} a_{15}^*\right|^2 \\
&+ 2\left|a_0 a_6^* + a_1 a_7^* + a_8 a_{14}^* + a_9 a_{15}^*\right|^2 + 2\left|a_2 a_6^* + a_3 a_7^* + a_{10} a_{14}^* + a_{11} a_{15}^*\right|^2 \\
&+ 2\left|a_0 a_4^* + a_2 a_6^* + a_8 a_{12}^* + a_{10} a_{14}^*\right|^2 + 2\left|a_0 a_1^* + a_2 a_3^* + a_8 a_9^* + a_{10} a_{11}^*\right|^2 \\
&+ 2\left|a_1 a_4^* + a_3 a_6^* + a_9 a_{12}^* + a_{11} a_{14}^*\right|^2 + 2\left|a_4 a_5^* + a_6 a_7^* + a_{12} a_{13}^* + a_{14} a_{15}^*\right|^2 \\
&+ 2\left|a_0 a_5^* + a_2 a_7^* + a_8 a_{13}^* + a_{10} a_{15}^*\right|^2 + 2\left|a_1 a_5^* + a_3 a_7^* + a_9 a_{13}^* + a_{11} a_{15}^*\right|^2 \\
&+ 2\left|a_0 a_2^* + a_4 a_6^* + a_8 a_{10}^* + a_{12} a_{14}^*\right|^2 + 2\left|a_0 a_1^* + a_4 a_5^* + a_8 a_9^* + a_{12} a_{13}^*\right|^2 \\
&+ 2\left|a_1 a_2^* + a_5 a_6^* + a_9 a_{10}^* + a_{13} a_{14}^*\right|^2 + 2\left|a_2 a_3^* + a_6 a_7^* + a_{10} a_{11}^* + a_{14} a_{15}^*\right|^2 \\
&+ 2\left|a_0 a_3^* + a_4 a_7^* + a_8 a_{11}^* + a_{12} a_{15}^*\right|^2 + 2\left|a_1 a_3^* + a_5 a_7^* + a_9 a_{11}^* + a_{13} a_{15}^*\right|^2
\end{aligned}$$